\documentclass[graybox]{svmult}

\usepackage{mathptmx}       
\usepackage{helvet}         
\usepackage{courier}        
\usepackage{type1cm}        
\usepackage{makeidx}         
\usepackage{graphicx}        
\usepackage{multicol}        
\usepackage[bottom]{footmisc}

\newcommand\rf[1]{(\ref{eq:#1})}
\newcommand\lab[1]{\label{eq:#1}}
\newcommand\nonu{\nonumber}
\newcommand\br{\begin{eqnarray}}
\newcommand\er{\end{eqnarray}}
\newcommand\be{\begin{equation}}
\newcommand\ee{\end{equation}}

\newcommand\lb{\lbrack}
\newcommand\rb{\rbrack}
\newcommand\llangle{\left\langle}
\newcommand\rrangle{\right\rangle}

\newcommand\llb{\left\lbrack}
\newcommand\rrb{\right\rbrack}

\renewcommand\({\left(}
\renewcommand\){\right)}
\newcommand\bv{\bigm\vert}               

\newcommand\bc{\begin{center}}
\newcommand\ec{\end{center}}

\newcommand\partder[2]{\frac{{\partial {#1}}}{{\partial {#2}}}}

\renewcommand\b{\beta}

\renewcommand\d{\delta}
\newcommand\eps{\epsilon}
\newcommand\vareps{\varepsilon}
\newcommand\g{\gamma}
\newcommand\G{\Gamma}

\newcommand\h{\frac{1}{2}}
\renewcommand\k{\kappa}
\renewcommand\l{\lambda}
\renewcommand\L{\Lambda}
\newcommand\m{\mu}
\newcommand\n{\nu}

\newcommand\pa{\partial}

\newcommand\s{\sigma}

\renewcommand\t{\tau}
\renewcommand\th{\theta}

\newcommand\cE{{\mathcal E}}
\newcommand\cF{{\mathcal F}}

\newcommand\cJ{{\mathcal J}}

\newcommand\cV{{\mathcal V}}

\newcommand{\ct}[1]{\cite{#1}}
\newcommand{\bib}[1]{\bibitem{#1}}

\newcommand\PRL[3]{\textsl{Phys. Rev. Lett.} \textbf{#1} (#2) #3}

\newcommand\PRD[3]{\textsl{Phys. Rev.} \textbf{D#1} (#2) #3}

\newcommand\PLB[3]{\textsl{Phys. Lett.} \textbf{#1B} (#2) #3}
\newcommand\CQG[3]{\textsl{Class. Quantum Grav.} \textbf{#1} (#2) #3}

\newcommand\AoP[3]{\textsl{Ann. of Phys.} \textbf{#1} (#2) #3}

\newcommand\IJMPA[3]{\textsl{Int. J. Mod. Phys.} \textbf{A#1} (#2) #3}

\newcommand\MPLA[3]{\textsl{Mod. Phys. Lett.} \textbf{A#1} (#2) #3}

\newcommand\xdot{\stackrel{.}{x}}

\newcommand\etadot{\stackrel{.}{\eta}}

\begin{document}

\title*{Lightlike Braneworlds in Anti-de Sitter Bulk Space-times}
\author{Eduardo Guendelman, Alexander Kaganovich, Emil Nissimov, Svetlana Pacheva}
\institute{Eduardo Guendelman and Alexander Kaganovich 
\at Department of Physics, Ben-Gurion University of the Negev, Beer-Sheva, Israel,\\
\email{guendel@bgu.ac.il}, \email{alexk@bgu.ac.il}
\and Emil Nissimov and Svetlana Pacheva 
\at Institute for Nuclear Research and Nuclear Energy, Bulgarian Academy of Sciences, 
Sofia, Bulgaria,
\email{nissimov@inrne.bas.bg}, \email{svetlana@inrne.bas.bg}}
\maketitle

\abstract{We consider five-dimensional Einstein-Maxwell-Kalb-Ramond system
self-consistently coupled to a {\em lightlike} 3-brane, where the latter acts
as material, charge and variable cosmological constant source. We find wormhole-like 
solutions 
whose total space-time manifold consists of either (a)
two ``universes'', which are identical copies of the exterior space-time region 
(beyond the horizon) of 5-dimensional Schwarzschild-anti-de Sitter black hole,
or (b) a ``right'' ``universe'' comprising the exterior space-time region of
Reissner-Nordstr{\"o}m-anti-de Sitter black hole and a ``left'' ``universe'' being 
the Rindler ``wedge'' of 5-dimensional flat Minkowski space. The wormhole
``throat'' connecting these ``universes'', which is located on their common
horizons, is self-consistently occupied by the lightlike 3-brane as a direct result
of its dynamics given by an explicit reparametrization-invariant world-volume 
Lagrangian action. The intrinsic world-volume metric on the 3-brane turns
out to be flat, which allows its interpretation as a {\em lightlike}
braneworld.}


\section{Introduction}
\label{sec:1}
Lightlike branes (\textsl{``LL-branes''} for short) play an important role
in modern general relativity. \textsl{LL-branes} are singular null (lightlike) 
hypersurfaces in Riemannian space-time which provide dynamical description of 
various physically important  phenomena in cosmology and astrophysics such as:
(i) impulsive lightlike signals arising in cataclysmic astrophysical events 
(supernovae, neutron star collisions) \ct{barrabes-hogan}; 
(ii) dynamics of horizons in black hole physics -- the so called ``membrane paradigm''
\ct{membrane-paradigm};
(iii) the thin-wall approach to domain walls coupled to gravity 
\ct{Israel-66}--\ct{Berezin-etal}.
More recently, \textsl{LL-branes} became significant also in the context of
modern non-perturbative string theory 
\ct{kogan-01,mateos-02,nonperturb-string-1,nonperturb-string-2}

In our previous papers \ct{LL-main-1}--\ct{beograd-2010} we have 
provided an explicit reparametrization invariant world-volume Lagrangian formulation of
lightlike $p$-branes (a brief review is given in Section 2)
and we have used them to construct various types of 
wormhole, 
regular black hole and lightlike braneworld solutions in $D\!=\!4$ or 
higher-dimensional asymptotically flat or asymptotically anti-de Sitter bulk 
space-times (for a detailed account of the general theory of wormholes see
the book \ct{visser-book} and also refs.\ct{visser-1}--\ct{bronnikov-2}). 
In particular, in refs.\ct{BR-WH-1,BR-WH-2,beograd-2010} we have
shown that lightlike branes can trigger a series of spontaneous
compactification-decompactification transitions of space-time regions,
\textsl{e.g.}, from ordinary compactified (``tube-like'') 
Levi-Civita-Bertotti-Robinson \ct{LC-BR-1,LC-BR-2,LC-BR-3}
space to non-compact Reissner-Nordstr{\"o}m or 
Reissner-Nordstr{\"o}m-de-Sitter region or {\sl vice versa}. 
Let us note that wormholes with ``tube-like'' structure (and regular black holes with 
``tube-like'' core) have been previously obtained within different contexts in 
refs.\ct{eduardo-wh}--\ct{zaslavskii-3}.

Let us emphasize the following characteristic features of \textsl{LL-branes} which 
drastically distinguish them from ordinary Nambu-Goto branes: 

(i) They describe intrinsically lightlike modes, whereas Nambu-Goto branes describe
massive ones.

(ii) The tension of the \textsl{LL-brane} arises as an {\em additional
dynamical degree of freedom}, whereas Nambu-Goto brane tension is a given
{\em ad hoc} constant. 
The latter characteristic feature significantly distinguishes our \textsl{LL-brane}
models from the previously proposed {\em tensionless} $p$-branes (for a
review of the latter, see Ref.~\ct{lindstroem-etal}) which rather resemble 
a $p$-dimensional continuous distribution of massless point-particles. 

(iii) Consistency of \textsl{LL-brane} dynamics in a spherically or axially
symmetric gravitational background of codimension one requires the presence
of a horizon which is automatically occupied by the \textsl{LL-brane}
(``horizon straddling'' according to the terminology of 
ref.~\ct{Barrabes-Israel}).

(iv) When the \textsl{LL-brane} moves as a {\em test} brane in spherically or 
axially symmetric gravitational backgrounds its dynamical tension exhibits 
exponential ``inflation/deflation'' time behavior \ct{inflation-all-1}
-- an effect similar to the ``mass inflation'' effect around black hole horizons
\ct{israel-poisson-1,israel-poisson-2}. 

Here we will focus on studying $4$-dimensional lightlike braneworlds
in 5-dimensional bulk anti-de Sitter spaces -- an alternative to the standard 
Randall-Sundrum scenario \ct{randall-sundrum-1,randall-sundrum-2} 
(for a systematic overview to braneworld theory, see 
\ct{mannheim,gregory,maartens-koyama}). Namely, we will present explicit 
solutions of 5-dimensionals Einstein-Maxwell\--Kalb-Ramond system self-consistently
interacting with codimension-one \textsl{LL-branes}, which are special kinds
of {\em ``wormhole''-like} space-times of either one of the following structures:

\vspace{.1in}
(A) two ``universes'' which are identical copies of the exterior space-time region 
(beyond the horizon) of 5-dimensional 
Schwarzschild-anti-de Sitter black hole;

(B) ``right'' ``universe'' comprising the exterior space-time region of
Reissner-Nordstr{\"o}m-anti-de Sitter black hole and ``left'' ``universe'' being 
the Rindler ``wedge'' of 5-dimensional flat Minkowski space.

\vspace{.1in} \noindent
Both ``right'' and ``left'' ``universes'' in (A)-(B) are glued together along their 
common horizons occupied by the \textsl{LL-brane} with {\em flat} 4-dimensional 
intrinsic world-volume metric, in other words, a flat lightlike braneworld
(\textsl{LL-braneworld}) at the wormhole ``throat''. In case (A) the 
\textsl{LL-brane} is electrically neutral whereas in case (B) it is both 
electrically charged as well as it couples also to a bulk Kalb-Ramond tensor gauge field.

\section{Lagrangian Formulation of Lightlike Brane Dynamics}
\label{sec:2}
In what follows we will consider gravity/gauge-field system
self-consistently interacting with a lightlike $p$-brane 
of codimension one ($D=(p+1)+1$). In a series of previous papers
\ct{LL-main-1}--\ct{beograd-2010}
we have proposed manifestly reparametrization invariant world-volume Lagrangian 
formulation in several dynamically equivalent forms of \textsl{LL-branes} coupled to 
bulk gravity $G_{\m\n}$ and bulk gauge fields, in particular, 
Maxwell $A_\m$ and Kalb-Ramond $A_{\m_1\ldots\m_{D-1}}$. 
Here we will use our Polyakov-type formulation given by the world-volume action:
\br
S_{\rm LL}\lb q,\b\rb  = - \h \int d^{p+1}\s\, T b_0^{\frac{p-1}{2}}\sqrt{-\g}
\llb \g^{ab} {\bar g}_{ab} - b_0 (p-1)\rrb \; ,
\lab{LL-action+EM+KR} \\
- \frac{\b}{(p+1)!} \int d^{p+1}\s\,\vareps^{a_1\ldots a_{p+1}}
\pa_{a_1} X^{\m_1}\ldots\pa_{a_{p+1}} X^{\m_{p+1}} A_{\m_1\ldots\m_{p+1}}
\er
where:
\be
{\bar g}_{ab} \equiv \pa_a X^\m G_{\m\n} \pa_b X^\n 
- \frac{1}{T^2} (\pa_a u + qA_a)(\pa_b u  + qA_b) 
\quad , \quad A_a \equiv \pa_a X^\m A_\m \; .
\lab{ind-metric-ext-A}
\ee
Here and below the following notations are used:
\begin{itemize}
\item
$X^\m (\s)$ are the $p$-brane embedding coordinates in the bulk
$D$-dimensional space-time with Riemannian metric
$G_{\m\n}(x)$ ($\m,\n = 0,1,\ldots ,D-1$); 
$(\s)\equiv \(\s^0 \equiv \t,\s^i\)$ with $i=1,\ldots ,p$;
$\pa_a \equiv \partder{}{\s^a}$.
\item
$\g_{ab}$ is the {\em intrinsic} Riemannian metric on the world-volume with
$\g = \det \Vert\g_{ab}\Vert$;
$g_{ab}$ is the {\em induced} metric on the world-volume:
\be
g_{ab} \equiv \pa_a X^{\m} G_{\m\n}(X) \pa_b X^{\n} \; ,
\lab{ind-metric}
\ee
which becomes {\em singular} on-shell (manifestation of the lightlike nature), 
\textsl{cf.} Eq.\rf{on-shell-singular-A} below); 
$b_0$ is a positive constant measuring the world-volume ``cosmological constant''.
\item
$u$ is auxiliary world-volume scalar field defining the lightlike direction
of the induced metric 
(see Eq.\rf{on-shell-singular-A} below) and it is a non-propagating degree of freedom
(cf. ref.\ct{beograd-2010}).
\item
$T$ is {\em dynamical (variable)} brane tension (also a non-propagating
degree of freedom). 
\item
The coupling parameters $q$ and $\b$ are the electric surface charge density
and the Kalb-Rammond charge of the \textsl{LL-brane}, respectively.
\end{itemize}

The corresponding equations of motion w.r.t. $X^\m$, $u$, $\g_{ab}$ and $T$ read accordingly
(using short-hand notation \rf{ind-metric-ext-A}):
\br
\pa_a \( T \sqrt{|{\bar g}|} {\bar g}^{ab}\pa_b X^\m\)
+ T \sqrt{|{\bar g}|} {\bar g}^{ab} \pa_a X^\l \pa_b X^\n \G^\m_{\l\n}
\nonu \\
+ \frac{q}{T} \sqrt{|{\bar g}|} {\bar g}^{ab}
\pa_a X^\n (\pa_b u  + qA_b) F_{\l\n}G^{\m\l} 
\nonu \\
- \frac{\b}{(p+1)!} \vareps^{a_1\ldots a_{p+1}} \pa_{a_1} X^{\m_1} \ldots
\pa_{a_{p+1}} X^{\m_{p+1}} F_{\l\m_1\dots\m_{p+1}} G^{\l\m} = 0 \; ,
\lab{X-eqs-NG-A} \\
\pa_a \(\frac{1}{T} \sqrt{|{\bar g}|} {\bar g}^{ab}(\pa_b u  + qA_b)\) = 0
\quad ,\quad  
\g_{ab} = \frac{1}{b_0} {\bar g}_{ab}  \; ,
\lab{u-gamma-eqs-NG-A} \\
T^2 + \eps {\bar g}^{ab}(\pa_a u  + qA_a)(\pa_b u  + qA_b) = 0 \; .
\lab{T-eq-NG-A}
\er
Here ${\bar g} = \det\Vert {\bar g}_{ab} \Vert$, $\G^\m_{\l\n}$ denotes the 
Christoffel connection for the bulk metric $G_{\m\n}$ and:
\be
F_{\m\n} = \pa_\m A_\n - \pa_\n A_\m \quad ,\quad
F_{\m_1\ldots\m_D} = D\pa_{[\m_1} A_{\m_2\ldots\m_D]} =
\cF \sqrt{-G} \vareps_{\m_1\ldots\m_D}
\lab{F-KR}
\ee
are the corresponding gauge field strengths.

The on-shell singularity of the induced metric $g_{ab}$ \rf{ind-metric}, \textsl{i.e.}, 
the lightlike property, directly follows Eq.\rf{T-eq-NG-A} and the definition of
${\bar g}_{ab}$ \rf{ind-metric-ext-A}:
\be
g_{ab} \({\bar g}^{bc}(\pa_c u  + qA_c)\) = 0 \; .
\lab{on-shell-singular-A}
\ee

Explicit world-volume reparametrization invariance of the \textsl{LL-brane} action
\rf{LL-action+EM+KR} allows to introduce the standard synchronous gauge-fixing conditions
for the intrinsic world-volume metric  
\be
\g^{00} = -1 \quad ,\quad \g^{0i} = 0 \;\; (i=1,\ldots,p) \; .
\lab{gauge-fix}
\ee
which reduces Eqs.\rf{u-gamma-eqs-NG-A}--\rf{T-eq-NG-A} to the following relations:
\br
\frac{(\pa_0 u + qA_0)^2}{T^2} = b_0 + g_{00} \quad ,\quad 
\pa_i u + qA_i= (\pa_0 u + qA_0) g_{0i} \( b_0 + g_{00}\)^{-1} \; ,
\nonu \\
g_{00} = g^{ij} g_{0i} g_{0j} \;\; ,\;\;
\pa_0 \(\sqrt{g^{(p)}}\) + \pa_i \(\sqrt{g^{(p)}}g^{ij} g_{0j}\) = 0 \;\; ,\;\; 
g^{(p)} \equiv \det\Vert g_{ij}\Vert \; ,
\lab{g-rel}
\er
(recall that $g_{00},g_{0i},g_{ij}$ are the components of the induced metric 
\rf{ind-metric}; $g^{ij}$ is the inverse matrix of $g_{ij}$).
Then, as shown in refs.\ct{LL-main-1}--\ct{beograd-2010}, 
consistency of \textsl{LL-brane} dynamics in static
``spherically-symmetric''-type backgrounds (in what follows we will use 
Eddington-Finkelstein coordinates, $dt=dv-\frac{d\eta}{A(\eta)}$):
\br
ds^2 = - A(\eta) dv^2 + 2dv d\eta + C(\eta) h_{ij}(\th) d\th^i d\th^j \; ,
\nonu \\
F_{v\eta} = F_{v\eta} (\eta)\; ,\; \mathrm{rest}=0 \quad ,\quad \cF = \cF (\eta) \; ,
\lab{static-spherical-EF}
\er
with the standard embedding ansatz:
\be
X^0\equiv v = \t \quad, \quad X^1\equiv \eta = \eta (\t) \quad, \quad 
X^i\equiv \th^i = \s^i \;\; (i=1,\ldots ,p) \; .
\lab{X-embed}
\ee
requires the corresponding background \rf{static-spherical-EF} to possess a horizon 
at some $\eta\!=\!\eta_0$, which is automatically occupied by the \textsl{LL-brane}.

Indeed, in the case of \rf{static-spherical-EF}--\rf{X-embed} Eqs.\rf{g-rel}
reduce to:
\be
g_{00} = 0 \;\; ,\;\; 
\pa_0 C \bigl(\eta (\t)\bigr) \equiv \etadot \pa_\eta C\bv_{\eta = \eta (\t)} = 0 
\;\; ,\;\;
\frac{(\pa_0 u + qA_0)^2}{T^2} = b_0 \;\; ,\;\; \pa_i u =0 
\lab{g-rel-0}
\ee
($\etadot \equiv \pa_0 \eta \equiv \pa_\t \eta (\t)$).
Thus, in the generic case of non-trivial dependence of $C(\eta)$ on the
``radial-like'' coordinate $\eta$, the first two relations in \rf{g-rel-0} yield:
\be
\etadot = \h A \bigl(\eta (\t)\bigr) \quad ,\quad \etadot = 0 \quad \to \quad
\eta (\t) = \eta_0 = \mathrm{const} \quad ,\quad A(\eta_0) = 0 \; .
\lab{straddling}
\ee
The latter property is called ``horizon straddling'' according to the terminology
of ref.\ct{Barrabes-Israel}. Similar ``horizon straddling'' has been found also for 
\textsl{LL-branes} moving in rotating axially symmetric (Kerr or Kerr-Newman) and 
rotating cylindrically symmetric black hole backgrounds 
\ct{Kerr-rot-WH-1,Kerr-rot-WH-2}. 

\section{Gravity/Gauge-Field System Interacting with Lightlike Brane}
\label{sec:3}

The generally covariant and manifestly world-volume reparametrization-invariant 
Lagrangian action describing a bulk Einstein-Maxwell-Kalb-Ramond system
(with bulk cosmological constant $\L$)
self-consistently interacting with a codimension-one \textsl{LL-brane} is given by:
\be
S = \int\!\! d^D x\,\sqrt{-G}\,\llb \frac{R(G)-2\L}{16\pi} 
- \frac{1}{4} F_{\m\n}F^{\m\n} 
- \frac{1}{D! 2} F_{\m_1\ldots\m_D}F^{\m_1\ldots\m_D}\rrb 
+ S_{\rm LL}\lb q,\b\rb \; ,
\lab{E-M-KR+LL}
\ee
where again $F_{\m\n}$ and $F_{\m_1\ldots\m_D}$ are the Maxwell and Kalb-Ramond
field-strengths \rf{F-KR} and $S_{\rm LL}\lb q,\b\rb$ indicates the world-volume action
of the \textsl{LL-brane} of the form \rf{LL-action+EM+KR}. It is now the 
\textsl{LL-brane} which will be the material and charge source for gravity and
electromagnetism, as well as it will generate dynamically an additional
space-varying bulk cosmological constant (see Eq.\rf{T-EM-KR} and 
second relation \rf{eqsys-4} below).

The equations of motion resulting from \rf{E-M-KR+LL} read:

(a) Einstein equations:
\be
R_{\m\n} - \h G_{\m\n} R + \L G_{\m\n} =
8\pi \( T^{(EM)}_{\m\n} + T^{(KR)}_{\m\n} + T^{(\mathrm{brane})}_{\m\n}\) \; ;
\lab{Einstein-eqs}
\ee

(b) Maxwell equations:
\be
\pa_\n \Bigl\lb\sqrt{-G} F_{\k\l} G^{\m\k} G^{\n\l}\Bigr\rb 
+ j_{(\mathrm{brane})}^\m = 0 \; ;
\lab{Maxwell-eqs}
\ee

(c) Kalb-Ramond equations (recall definition of $\cF$ in \rf{F-KR}):
\be
\vareps^{\n\m_1\ldots\m_{p+1}} \pa_\n \cF 
- J_{(\mathrm{brane})}^{\m_1\ldots\m_{p+1}} = 0 \; ;
\lab{F-KR-eqs}
\ee

(d) The \textsl{LL-brane} equations of motion have already been written down in
\rf{X-eqs-NG-A}--\rf{T-eq-NG-A} above.

The energy-momentum tensors of bulk gauge fields are given by:
\be
T^{(EM)}_{\m\n} = F_{\m\k}F^{\m\n} - G_{\m\n}\frac{1}{4}F_{\k\l}F^{\k\l} 
\quad ,\quad T^{(KR)}_{\m\n} = - \h \cF^2 G_{\m\n} \; , 
\lab{T-EM-KR}
\ee
where the last relation indicates that $\L \equiv 4\pi \cF^2$ can be
interpreted as dynamically generated cosmological ``constant''.

The energy-momentum (stress-energy) tensor $T^{(\mathrm{brane})}_{\m\n}$ and the 
electromagnetic $ j_{(\mathrm{brane})}^\m$  and Kalb-Ramond
$J_{(\mathrm{brane})}^{\m_1\ldots\m_{p+1}}$ charge current densities 
of the \textsl{LL-brane} arestraightforwardly derived from the pertinent 
\textsl{LL-brane} action \rf{LL-action+EM+KR}:
\br
T_{(\mathrm{brane})}^{\m\n} = 
- \int\!\! d^{p+1}\s\,\frac{\d^{(D)}\Bigl(x-X(\s)\Bigr)}{\sqrt{-G}}
\, T\,\sqrt{|{\bar g}|} {\bar g}^{ab} \pa_a X^\m \pa_b X^\n \; ,
\lab{T-brane-A} \\
j_{(\mathrm{brane})}^\m = - q \int\!\! d^{p+1}\s\,\d^{(D)}\Bigl(x-X(\s)\Bigr)
\sqrt{|{\bar g}|} {\bar g}^{ab}\pa_a X^\m \(\pa_b u + qA_b\)T^{-1} \; ,
\lab{j-brane-A} \\
J_{(\mathrm{brane})}^{\m_1\ldots\m_{p+1}} = 
\b \int\!\! d^{p+1}\s\,\d^{(D)}\Bigl(x-X(\s)\Bigr) \vareps^{a_1\ldots a_{p+1}} 
\pa_{a_1} X^{\m_1}\ldots \pa_{a_{p+1}} X^{\m_{p+1}} \; .
\lab{j-KR-brane-A}
\er


Construction of ``wormhole''-like solutions of static ``spherically-symmetric''-type
\rf{static-spherical-EF} for the coupled gravity-gauge-field-\textsl{LL-brane} 
system \rf{E-M-KR+LL} proceeds along the following simple steps:

(i) Choose ``vacuum'' static ``spherically-symmetric''-type solutions
\rf{static-spherical-EF} of \rf{Einstein-eqs}--\rf{F-KR-eqs} (\textsl{i.e.}, 
without the delta-function terms due to the \textsl{LL-branes}) in each region 
$-\infty < \eta < \eta_0$ and $\eta_0 <\eta < \infty$ with a common horizon
at $\eta=\eta_0$;

(ii) The \textsl{LL-brane} automatically locates itself on the horizon
according to ``horizon straddling'' property \rf{straddling};

(iii) Match the discontinuities of the derivatives of the metric and the gauge field
strength \rf{static-spherical-EF} across the horizon at $\eta = \eta_0$ using the 
explicit expressions for the \textsl{LL-brane} stress-energy tensor, electromagnetic
and Kalb-Ramond charge current densities \rf{T-brane-A}--\rf{j-KR-brane-A}.

Using \rf{g-rel}--\rf{X-embed} we find:
\br
T_{(\mathrm{brane})}^{\m\n} = S^{\m\n}\,\d (\eta-\eta_0) \quad ,\quad
j_{(\mathrm{brane})}^\m = \d^\m_0 q\sqrt{\det\Vert G_{ij}\Vert}\, \d (\eta-\eta_0) \; ,
\lab{T-j-0} \\
\frac{1}{(p+1)!} \vareps_{\m\n_1\ldots\n_{p+1}} J^{\n_1\ldots\n_{p+1}} = 
\b\,\d^\eta_\m \d\(\eta - \eta_0 \)
\nonu
\er
where $G_{ij} = C(\eta) h_{ij}(\th)$ (\textsl{cf.} \rf{static-spherical-EF})
and the surface energy-momentum tensor reads:
\be
S^{\m\n} \equiv \frac{T}{b_0^{1/2}}\, 
\( \pa_\t X^\m \pa_\t X^\n - b_0 G^{ij} \pa_i X^\m \pa_j X^\n 
\)_{v=\t,\,\eta=\eta_0,\,\th^i =\s^i} \; .
\lab{T-S-brane} 
\ee
The non-zero components of $S_{\m\n}$ (with lower indices) and its trace are:
\be
S_{\eta\eta} = \frac{T}{b_0^{1/2}} \quad ,\quad 
S_{ij} = - T b_0^{1/2} G_{ij} \quad ,\quad S^\l_\l = - pTb_0^{1/2} \; .
\lab{S-comp}
\ee
Taking into account \rf{T-j-0}--\rf{S-comp} together with
\rf{static-spherical-EF}--\rf{straddling}, the matching relations at the
horizon $\eta =\eta_0$ become \ct{BR-WH-1,BR-WH-2,beograd-2010}
(for a systematic introduction to the formalism of matching different bulk
space-time geometries on codimension-one hypersurfaces (``thin shells'') see
the textbook \ct{poisson-kit}):

(i) Matching relations from Einstein eqs.\rf{Einstein-eqs}:
\be
\llb \pa_\eta A \rrb_{\eta_0} = - 16\pi T \sqrt{b_0} 
\quad,\quad 
\llb \pa_\eta \ln C \rrb_{\eta_0} = - \frac{16\pi}{p\sqrt{b_0}} T
\lab{eqsys-1-2}
\ee
with notation $\bigl\lb Y \bigr\rb_{\eta_0} \equiv 
Y\bv_{\eta \to \eta_0 +0} - Y\bv_{\eta \to \eta_0 -0}$ for any quantity $Y$.

(ii) Matching relation from gauge field eqs.\rf{Maxwell-eqs}--\rf{F-KR-eqs}:
\be
\llb F_{v\eta} \rrb_{\eta_0} = q \quad ,\quad
\llb \cF \rrb_{\eta_0} = -\b \; .
\lab{eqsys-4}
\ee

(iii) $X^0$-equation of motion of the \textsl{LL-brane} (the only non-trivial
contribution of second-order \textsl{LL-brane} eqs.\rf{X-eqs-NG-A} in the
case of embedding \rf{X-embed}):
\be
\frac{T}{2} \( \llangle \pa_\eta A \rrangle_{\eta_0} 
+ p b_0 \llangle \pa_\eta \ln C \rrangle_{\eta_0} \)
-  \sqrt{b_0} \( q \llangle F_{v\eta}\rrangle_{\eta_0}
- \b \llangle \cF\rrangle_{\eta=\eta_0}\) = 0
\lab{eqsys-3}
\ee
with notation $\llangle Y \rrangle_{\eta_0} \equiv 
\h \( Y\bv_{\eta \to \eta_0 +0} + Y\bv_{\eta \to \eta_0 -0}\)$. 

\section{Explicit Solutions: Braneworlds via Lightlike Brane}
\label{sec:4}

Consider 5-dimensional AdS-Schwarzschild black hole in Eddington-Finkelstein
coordinates $(v,r,\vec{x})$ (with $\vec{x}\equiv (x^1,x^2,x^3)$):
\be
ds^2 = - A(r) dv^2 + 2dv\,dr + K r^2 d\vec{x}^2 \quad,\quad A(r) = Kr^2 - m/r^2 \; ,
\lab{AdS-Schw-BH}
\ee
where $\L =-6K$ is the bare negative 5-dimensional cosmological constant
and $m$ is the mass parameter of the black hole. The pertinent horizon is
located at:
\be
A(r_0) = 0 \to r_0 = \( m/K\)^{1/4}\;\; ,\;\; \mathrm{where} \;\;
\pa_r A (r_0) >0 \; .
\lab{AdS-Schw-BH-hor}
\ee

First, let us consider self-consistent Einstein-\textsl{LL-brane} system \rf{E-M-KR+LL}
with a neutral \textsl{LL-brane} source (\textsl{i.e.} no \textsl{LL-brane} couplings to bulk 
Maxwell and Kalb-Ramond gauge fields: $q,\b = 0$ in $S_{\rm LL}\lb q,\b\rb$).
A simple trick to obtain {\em ``wormhole''}-like solution to this coupled system
is to change variables in \rf{AdS-Schw-BH}:
\be
r \to r(\eta) = r_0 + |\eta|
\lab{var-change}
\ee
with $r_0$ being the AdS-Schwarzschild horizon \rf{AdS-Schw-BH-hor}, 
where now $\eta \in (-\infty,+\infty)$, \textsl{i.e.}, consider:
\br
ds^2 = - A(\eta) dv^2 + 2dv\,d\eta + C(\eta) d\vec{x}^2 \; ,
\lab{AdS5-BW-1}\\
A(\eta) = K (r_0 + |\eta|)^2 - \frac{m}{(r_0 + |\eta|)^2} \quad ,\quad
C(\eta) = K (r_0 + |\eta|)^2 \; ,
\lab{AdS5-BW-2}\\
A(0) = 0 \quad ,\quad A(\eta) >0 \; \mathrm{for}\; \eta \neq 0  \; .
\nonu
\er
Obviously, \rf{var-change} is {\em not} a smooth local coordinate transformation 
due to $|\eta|$. The coefficients of the new metric \rf{AdS5-BW-1}--\rf{AdS5-BW-2} 
are continuous at the horizon $\eta_0 = 0$ with discontinuous first derivates
across the horizon. The \textsl{LL-brane} automatically locates itself on
the horizon according to the ``horizon-straddling'' property of its
world-volume dynamics \rf{straddling}. 

Substituting \rf{AdS5-BW-1}--\rf{AdS5-BW-2} into the matching relations 
\rf{eqsys-1-2}--\rf{eqsys-3} we find the following relation between bulk space-time 
parameters $(K= |\L|/6,m)$ and the \textsl{LL-brane} parameters $(T,b_0)$ : 
\be
T^2 = \frac{3}{8\pi^2}K \quad ,\quad T < 0 \quad ,\quad b_0 = \frac{2}{3}\sqrt{Km} \; .
\lab{AdS5-BW-3}
\ee
Taking into account second Eq.\rf{u-gamma-eqs-NG-A} and \rf{gauge-fix} 
the intrinsic metric $\g_{ab}$ on the \textsl{LL-braneworld} becomes {\em flat}:
\be
\g_{00} = -1 \quad ,\quad \g_{0i} = 0 \quad ,\quad \g_{ij} = \frac{3}{2} \d_{ij} \; .
\lab{flat-LLBW-1}
\ee

The solution \rf{AdS5-BW-2}--\rf{flat-LLBW-1} describes a ``wormhole''-like
$D=5$ bulk space-time consisting of two ``universes'' being identical copies of the exterior
region beyond the horizon ($r>r_0$) of the 5-dimensional AdS-Schwarzschild black hole
glued together along their common horizon (at $r=r_0$) by the
\textsl{LL-brane}, \textsl{i.e.}, the latter serving as a wormhole
``throat'', which in turn can be viewed as a \textsl{LL-braneworld} with flat
intrinsic geometry \rf{flat-LLBW-1}.

Let us now consider the 5-dimensional AdS-Reissner-Nordstr{\"o}m black hole 
(in Eddington-Finkelstein coordinates $(v,r,\vec{x})$):
\br
ds^2 = - A(r) dv^2 + 2dv\,dr + K r^2 d\vec{x}^2  \quad ,\quad \L =-6K \; ,
\nonu \\
A(r) = Kr^2 - \frac{m}{r^2} + \frac{Q}{r^4} \quad ,\quad
F_{vr} = \sqrt{\frac{3}{4\pi}} \frac{Q}{r^3} \; .
\lab{AdS-RN-BH}
\er
We can construct, following the same procedure, another {\em non-symmetric}
``wormhole''-like solution with a flat \textsl{LL-braneworld} occupying its
``throat'' provided the \textsl{LL-brane} is electrically charged and
couples to bulk Kalb-Ramond gauge field, \textsl{i.e.}, $q,\b \neq 0$ in 
\rf{E-M-KR+LL}, \rf{LL-action+EM+KR}. This solution describes:

(a) ``left'' universe being a 5-dimensional flat Rindler space-time -- the
Rindler ``wedge'' of $D=5$ Minkowski space \ct{rindler,MTW} (here $|\eta| = X^2$,
where $X$ is the standard Rindler coordinate):
\be
ds^2 = \eta dv^2 + 2dv\,d\eta + d\vec{x}^2 \quad , 
\quad  \; \mathrm{for}\;\eta <0 \; ;
\lab{Rindler-wedge}
\ee

(b) ``right'' universe comprizing the exterior $D=5$ space-time region of the
AdS-Reissner-Nordstr{\"o}m black hole beyond the {\em outer} 
AdS-Reissner-Nordstr{\"o}m horizon $r_0$ ($A(r_0)=0$ with $A(r)$ as in \rf{AdS-RN-BH} 
and where again we apply the non-smooth coordinate change \rf{var-change}):
\br
ds^2 = - A(\eta) dv^2 + 2dv\,d\eta + K (r_0 + \eta)^2 d\vec{x}^2
\lab{AdS-RN-1}\\
A(\eta) = K (r_0 + \eta)^2 - \frac{m}{(r_0 + \eta)^2} + \frac{Q^2}{(r_0 + \eta)^4} 
\lab{AdS-RN-2}\\
F_{v\eta} = \sqrt{\frac{3}{4\pi}} \frac{Q}{(r_0 + \eta)^3} \quad ,\quad
A(0) = 0 \;\; ,\;\; \pa_\eta A (0) >0 \;\; , \quad \mathrm{for}\; \eta > 0 \; .
\lab{AdS-RN-3}
\er

All physical parameters of the ``wormhole''-like solution 
\rf{Rindler-wedge}--\rf{AdS-RN-3} are determined in terms of $(q,\b)$ -- 
the electric and Kalb-Ramond \textsl{LL-braneworld} charges:
\br
m = \frac{3}{2\pi\b^2} \bigl( 1 + \frac{2q^2}{\b^2} \bigr) \quad ,\quad
Q^2 = \frac{9 q^2}{2\pi\b^6} \quad ,\quad |\L| \equiv 6K = 4\pi\b^2
\lab{AdS-RN-param}\\
|T| = \frac{1}{8\pi} \sqrt{\frac{3}{2}\sqrt{K} + 4\pi (\b^2 - q^2)} \; ,\;
b_0 = \frac{1}{6\sqrt{K}} \bigl\lb 1 + \frac{8\pi}{3}\sqrt{K} (\b^2 - q^2)\bigr\rb
\lab{LL-BW-param}
\er
Here again $T<0$. Let is stress the importance of the third relation in 
\rf{AdS-RN-param}. Namely, the dynamically generated space-varying effective cosmological
constant (cf. second Eq.\rf{T-EM-KR}) through the Kalb-Ramond coupling of the 
\textsl{LL-brane} (cf. second matching relation in \rf{eqsys-4})
has zero value in the ``right'' AdS-Reissner-Nordstr{\"o}m ``universe'' and
has positive value $4\pi\b^2$ in the ``left'' flat Rindler ``universe'' 
\rf{Rindler-wedge} compensating the negative bare cosmological constant $\L$.

The intrinsic metric $\g_{ab}$ on the \textsl{LL-braneworld} is again {\em flat}:
\be
\g_{00} = -1 \quad ,\quad \g_{0i} = 0 \quad ,\quad \g_{ij} = \frac{1}{b_0}\d_{ij}
\lab{flat-LLBW-2}
\ee

\section{Traversability and Trapping Near the Lightlike Braneworld}
\label{sec:5}

The ``wormhole''-like solutions presented in the previous Section share the
following important properties:

(a) The \textsl{LL-braneworlds} at the wormhole ``throats'' represent ``exotic''
matter with $T<0$, \textsl{i.e.}, negative brane tension implying 
violation of the null-energy conditions as predicted by general wormhole arguments
\ct{visser-book} (although the latter could be remedied via quantum fluctuations).

(b) The wormhole space-times constructed via \textsl{LL-branes} at
their ``throats'' are {\em not} traversable w.r.t. the ``laboratory'' time of a 
static observer in either of the different ``universes'' comprising the pertinent 
wormhole space-time manifold since the \textsl{LL-branes} sitting at the
``throats'' look as black hole horizons to the static observer. On the other hand, 
these wormholes {\em are traversable} w.r.t. the {\em proper time} of a 
traveling observer.

Indeed, proper-time traversability can be easily seen by considering dynamics of 
test particle of mass $m_0$ (``traveling observer'') in a wormhole background, 
which is described by the reparametrization-invariant world-line action:
\be
S_{\mathrm{particle}} = \h \int d\l \Bigl\lb\frac{1}{e}\xdot^\m \xdot^\n G_{\m\n}
- e m_0^2 \rb \; .
\lab{test-particle}
\ee
Using energy $\cE$ and orbital momentum $\cJ$ conservation and introducing the 
{\em proper} world-line time $s$ ($\frac{ds}{d\l}= e m_0$), the ``mass-shell''
constraint equation (the equation w.r.t. the ``einbein'' $e$) produced by the action 
\rf{test-particle}) yields:
\be
\(\frac{d\eta}{ds}\)^2 + \cV_{\mathrm{eff}} (\eta) = \frac{\cE^2}{m_0^2}
\quad ,\quad 
\cV_{\mathrm{eff}} (\eta) \equiv A(\eta) \Bigl( 1 + \frac{\cJ^2}{m_0^2 C(\eta)}\Bigr) 
\lab{particle-eq-2}
\ee
where the metric coefficients $A (\eta),\, C(\eta)$ are those in
\rf{static-spherical-EF}.

Since the ``effective potential'' $\cV_{\mathrm{eff}} (\eta)$ in \rf{particle-eq-2}
is everywhere non-negative and vanishes only at the wormhole throat(s)
($\eta = \eta_0$, where $A(\eta_0)=0$), ``radially'' moving test matter (\textsl{e.g.} 
a traveling observer) with zero ``impact'' parameter $\cJ=0$ and with sufficiently large 
energy $\cE$) will always cross from one ``universe'' to another
within {\em finite} amount of its proper-time (see Fig.1). Moreover, this test matter
(travelling observer) will ``shuttle'' between the turning points $\eta_{\pm}$:
\be
\cV_{\mathrm{eff}} (\eta_{\pm}) = \frac{\cE^2}{m_0^2} \quad ,\;\; 
\eta_{+} >0 \;\;, \;\; \eta_{-} < 0 \; ,
\lab{turning-points}
\ee
so that in fact it will be trapped in the vicinity of the \textsl{LL-braneworld}. 
This effect is analogous to the gravitational trapping of matter near domain
wall of a stable false vacuum bubble in cosmology \ct{eduardo-idan}.

\begin{figure}
\begin{center}
\includegraphics[width=8cm,keepaspectratio=true]{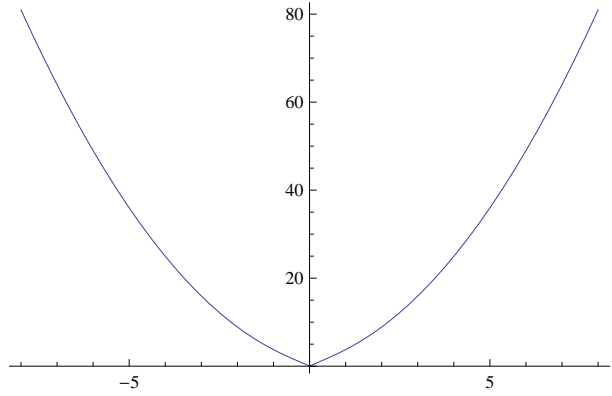}
\caption{Shape of the ``effective potential'' $\cV_{\mathrm{eff}} (\eta)=A(\eta)$
with $A(\eta)$ as in \rf{AdS5-BW-2}. Travelling observer along
the extra 5-th dimension will ``shuttle'' between the two 5-dimensional AdS ``universes''
crossing in either direction the 4-dimensional flat braneworld within {\em finite}
proper-time intervals.}
\end{center}
\end{figure}

\section{Discussion}
\label{sec:6}

Let us recapitulate the crucial properties of the dynamics of \textsl{LL-branes} 
interacting with gravity and bulk space-time gauge fields which enabled us to construct
the \textsl{LL-braneworld} solutions presented above:

\begin{itemize}
\item
(i) ``Horizon straddling'' -- automatic positioning of \textsl{LL-branes} on (one of) 
the horizon(s) of the bulk space-time geometry. 
\item
(ii) Intrinsic nature of the \textsl{LL-brane} tension as an additional 
{\em degree of freedom} unlike the case of standard Nambu-Goto $p$-branes;
(where it is a given \textsl{ad hoc} constant), and which might in particular acquire
negative values. Moreover, the variable tension feature significantly distinguishes 
\textsl{LL-brane} models from the previously proposed {\em tensionless} $p$-branes
-- the latter rather resemble $p$-dimensional continuous distributions of independent 
massless point-particles without cohesion among them.
\item
(iii) The stress-energy tensors of the \textsl{LL-branes} are systematically derived
from the underlying \textsl{LL-brane} Lagrangian actions and provide the appropriate
source terms on the r.h.s. of Einstein equations to enable the existence of
consistent non-trivial wormhole-like solutions. 
\item
(iv) Electrically charged \textsl{LL-branes} naturally produce {\em asymmetric}
wormholes with the \textsl{LL-branes} themselves materializing the wormhole ``throats''
and uniquely determining the pertinent wormhole parameters.
\item
(v) \textsl{LL-branes} naturally couple to Kalb-Ramond bulk space-time gauge fields 
which results in {\em dynamical} generation of space-time varying cosmological constant.
(vi) \textsl{LL-branes} naturally produce {\em lightlike} braneworlds
(extra dimensions are undetectable for observers confined on the
\textsl{LL-brane} universe).
\end{itemize}

\noindent
In our previous works we have also shown that:
\begin{itemize}
\item
(vii) \textsl{LL-branes} trigger sequences of spontaneous 
compactification/decompactification transitions of space-time 
\ct{BR-WH-1,BR-WH-2,beograd-2010}.
\item
(viii) \textsl{LL-branes} remove physical singularities of black holes \ct{reg-BH}.
\end{itemize}

The crucial importance of \textsl{LL-branes} in wormhole physics is
underscored by the role they are playing in the self-consistent contsruction
of the famous Einstein-Rosen ``bridge'' wormhole in its {\em original}
formulation \ct{einstein-rosen} -- historically the first explicit wormhole
solution. To this end let us make the following important remark. 
In several standard textbooks, \textsl{e.g.} \ct{MTW,Carroll}, 
the formulation of the Einstein-Rosen ``bridge'' uses the Kruskal-Szekeres manifold, 
where the Einstein-Rosen ``bridge'' geometry becomes {\em dynamical} 
(see ref.\ct{MTW}, p.839, Fig. 31.6, and ref.~\ct{Carroll}, p.228, Fig. 5.15). 
The latter notion of the Einstein-Rosen ``bridge'' is {\em not} equivalent to the 
original Einstein-Rosen's formulation in the classic paper \ct{einstein-rosen}, 
where the space-time manifold is {\em static} spherically
symmetric consisting of two identical copies of the outer Schwarzschild space-time region
($r>2m$) glued together
along the horizon at $r=2m$. Namely, the two regions in Kruskal-Szekeres space-time 
corresponding to the outer Schwarzschild space-time region ($r>2m$) and labeled 
$(I)$ and $(III)$ in ref.\ct{MTW} are generally {\em disconnected} and share only a 
two-sphere (the angular part) as a common border ($U=0, V=0$ in Kruskal-Szekeres coordinates),
whereas in the original Einstein-Rosen
``bridge'' construction \ct{einstein-rosen} the boundary between the two identical 
copies of the outer Schwarzschild space-time region ($r>2m$) is their common
horizon ($r=2m)$ -- a three-dimensional {\em lightlike} hypersurface. 
In refs.\ct{LL-main-4,Kerr-rot-WH-2} it has been shown that the Einstein-Rosen ``bridge'' 
in its original formulation
\ct{einstein-rosen} naturally arises as the simplest particular case of static 
spherically symmetric wormhole solutions produced by \textsl{LL-branes} as
gravitational sources, where the two identical ``universes'' with Schwarzschild
outer-region geometry are glued together by a \textsl{LL-brane} occupying
their common horizon -- the wormhole ``throat''. An understanding of this
picture within the framework of Kruskal-Szekeres manifold was subsequently
given in ref.\ct{poplawski}, which uses Rindler's identification of antipodal future 
event horizons.

One of the most interesting physical phenomena in wormhole physics is the 
well-known Misner-Wheeler ``charge without charge'' effect \ct{misner-wheeler}.
Namely, Misner and Wheeler have shown that wormholes connecting two asymptotically flat 
space-times provide the possibility of existence of electromagnetically
non-trivial solutions, where {\em without being produced by any charge source} the
flux of the electric field flows from one universe to the other, thus giving the impression 
of being positively charged in one universe and negatively charged in the other universe.

In our recent paper \ct{hiding} we found an opposite ``charge-hiding'' effect in wormhole
physics, namely, that a genuinely charged matter source of gravity and electromagnetism may 
appear {\em electrically neutral} to an external observer. This phenomenon takes place when 
coupling self-consistently an electrically charged \textsl{LL-brane} to gravity and
a {\em non-standard} form of nonlinear electrodynamics, whose Lagrangian contains a 
square-root of the ordinary Maxwell term: 
\be
L(F^2) = - \frac{1}{4} F^2 - \frac{f}{2} \sqrt{-F^2} \quad ,\quad
F^2 \equiv F_{\m\n} F^{\m\n} \; ,
\lab{GG}
\ee
$f$ being a positive coupling constant. In flat space-time the theory \rf{GG} is known 
to produce a QCD-like effective potential between charged fermions \ct{GG-1}--\ct{GG-6}. 
When coupled to gravity it generates an effective global cosmological constant 
$\L_{\mathrm{eff}}=2\pi f^2$ as well a nontrivial constant radial vacuum electric field 
$f/\sqrt{2}$ \ct{grav-cornell}. When in addition gravity and nonlinear
electrodynamics \rf{GG} also interact self-consistently with a charged \textsl{LL-brane}
we found in ref.\ct{hiding} a new type of wormhole solution which connects a non-compact 
``universe'', comprising the exterior region of Schwarzschild-de Sitter black hole beyond 
the internal (Schwarzschild-type horizon), to a Levi-Civita-Bertotti-Robinson-type 
``tube-like'' ``universe'' with two compactified dimensions 
(\textsl{cf.} \ct{LC-BR-1,LC-BR-2,LC-BR-3}) via a wormhole ``throat'' 
occupied by the charged \textsl{LL-brane}. In this solution the whole electric flux produced
by the charged \textsl{LL-brane} is pushed into the ``tube-like'' 
Levi-Civita-Bertotti-Robinson-type ``universe'' and thus the brane is detected as neutral by an
observer in the Schwarzschild-de-Sitter ``universe''.

In the subsequent recent paper \ct{hide-confine} we succeeded to find a truly 
``charge-confining'' wormhole solution when the coupled system of gravity and non-standard 
nonlinear electrodynamikcs \rf{GG} are self-consistently interacting with {\em two} separate 
oppositely charged \textsl{LL-branes}. Namely, we found a self-consistent ``two-throat'' 
wormhole solution where the ``left-most'' and the ``right-most'' ``universes''
are two identical copies of the exterior region of the electrically neutral 
Schwarzschild-de-Sitter black hole beyond the Schwarzschild horizon, whereas the ``middle''
``universe'' is of generalized Levi-Civita-Bertotti-Robinson ``tube-like'' form with 
geometry $dS_2 \times S^2$ ($dS_2$ is the two-dimensional de Sitter space). It comprises 
the finite-size intermediate region of $dS_2$ between its two horizons. Both ``throats'' 
are occupied by the two oppositely charged \textsl{LL-branes} and the whole
electric flux produced by the latter is confined entirely within the middle
finite-size ``tube-like'' ``universe''. 

One of the most important issues to be studied is the problem of stability
of the wormhole(-like) solutions with \textsl{LL-branes} at their
``troats'', in particular, the above presented \textsl{LL-braneworld}
solutions in anti-de Sitter bulk space-times. The ``horizon-straddling''
property \rf{straddling} of \textsl{LL-brane} dynamics will impose severe
restrictions on the impact of the perturbations of the bulk space-time geometry. 

\begin{acknowledgement}
E.N. and S.P. are grateful to the organizers for hospitality during the 
IX-th International Workshop ``Lie Theory and Its Applications in Physics''. 
E.N. and S.P. are supported by Bulgarian NSF grant \textsl{DO 02-257}.
Also, all of us acknowledge support of our collaboration through the exchange
agreement between the Ben-Gurion University 
and the Bulgarian Academy of Sciences.
\end{acknowledgement}
%
%


\begin{thebibliography}{99.}%
\bib{barrabes-hogan}
C. Barrab\'{e}s and P. Hogan, {\em ``Singular Null-Hypersurfaces in General
Relativity''} (World Scientific, Singapore, 2004).
%
\bib{membrane-paradigm}
K. Thorne, R. Price and D. Macdonald (Eds.), {\em ``Black Holes: The
Membrane Paradigm''} (Yale Univ. Press, New Haven, CT, 1986).
\bib{Israel-66}
W. Israel, \textsl{Nuovo Cim.} \textbf{B44}, 1 (1966); erratum, \textsl{Nuovo Cim.}
\textbf{B48}, 463 (1967). 
\bib{Barrabes-Israel}
C. Barrab\'{e}s and W. Israel, \PRD{43}{1991}{1129}. 
\bib{Dray-Hooft}
T. Dray and G. `t Hooft, \CQG{3}{1986}{825}. 
\bib{Berezin-etal}
V. Berezin, A. Kuzmin and I. Tkachev, \PRD{36}{1987}{2919}.
\bib{kogan-01}
I. Kogan and N. Reis, \IJMPA{16}{2001}{4567} ~(\textsl{arxiv:hep-th/0107163}). 
\bib{mateos-02}
D. Mateos and S. Ng, {\sl JHEP} {\bf 0208}, 005 (2002)
~(\textsl{arxiv:hep-th/0205291}). 
\bib{nonperturb-string-1}
D. Mateos, T. Mateos and P.K. Townsend,  \textsl{JHEP} {\bf 0312}, 017 (2003)
~(\textsl{arxiv:hep-th/0309114}). 
\bib{nonperturb-string-2}
A. Bredthauer, U. Lindstr{\"o}m, J. Persson and L. Wulff, \textsl{JHEP} 
{\bf 0402}, 051 (2004) ~(\textsl{arxiv:hep-th/0401159}).
\bib{LL-main-1}
E. Guendelman, A. Kaganovich, E. Nissimov and S. Pacheva, 
\PRD{72}{2005}{0806011} ~(\textsl{hep-th/0507193}). 
\bib{LL-main-2}
E. Guendelman, A. Kaganovich, E. Nissimov and S. Pacheva, 
\textsl{Fortschr. der Physik} {\bf 55}, 579 (2007) ~(\textsl{hep-th/0612091}). 
\bib{LL-main-3}
E. Guendelman, A. Kaganovich, E. Nissimov and S. Pacheva, 
\textsl{Fortschr. der Phys.} {\bf 57}, 566 (2009) 
~(\textsl{arxiv:0901.4443}[hep-th]). 
\bib{LL-main-4}
E. Guendelman, A. Kaganovich, E. Nissimov and S. Pacheva,
\PLB{681}{2009}{457}
~(\textsl{arxiv:0904.3198}[hep-th]). 
\bib{reg-BH}
E. Guendelman, A. Kaganovich, E. Nissimov and S. Pacheva,
\IJMPA{25}{2010}{1571-1596} ~(\textsl{arxiv:0908.4195}[hep-th]).
%
\bib{Kerr-rot-WH-1}
E. Guendelman, A. Kaganovich, E. Nissimov and S. Pacheva, 
\PLB{673}{2009}{288-292} ~(\textsl{arxiv:0811.2882}[hep-th]). 
\bib{Kerr-rot-WH-2}
E. Guendelman, A. Kaganovich, E. Nissimov and S. Pacheva,
\IJMPA{25}{2010}{1405} ~(\textsl{arxiv:0904.0401}[hep-th]).
%
\bib{BR-WH-1}
E. Guendelman, A. Kaganovich, E. Nissimov and S. Pacheva, 
\textsl{Gen. Rel. Grav.} {\bf 43}, 1487-1513 (2011) 
~(\textsl{arxiv:1007.4893}[hep-th]). 
\bib{BR-WH-2}
E. Guendelman, A. Kaganovich, E. Nissimov and S. Pacheva, in
special issue {\em ``Novel Ideas in Theoretical Physics and Astrophysics''} of
\textsl{Invertis Journal of Science and Technology}, to appear 
~(\textsl{arxiv:0912.3712}[hep-th]). 
\bib{beograd-2010}
E. Guendelman, A. Kaganovich, E. Nissimov and S. Pacheva, 
in {\sl ``Sixth Meeting in Modern Mathematical Physics''}, ed. by B. Dragovic and 
Z. Rakic, (Belgrade Inst. Phys. Press, Belgrade, 2011) ~(\textsl{arxiv:1011.6241}[hep-th]).
\bib{visser-book}
M. Visser, {\em ``Lorentzian Wormholes. From Einstein to Hawking''},
(Springer, Berlin, 1996).
\bib{visser-1}
Hochberg D, Visser M. \PRD{56}{1997}{4745}
~(\textsl{arxiv:9704082}[gr-qc]). 
\bib{visser-2}
Hochberg D, Visser M. \PRD{58}{1998}{044021}
~(\textsl{arxiv:9802046}[gr-qc]). 
\bib{WH-rev-1}
Lemos J, Lobo F, de Oliveira S. \PRD{68}{2003}{064004} 
~(\textsl{arxiv:0302049}[gr-qc]). 
\bib{WH-rev-2}
Sushkov S. \PRD{71}{2005}{043520} ~(\textsl{arxiv:0502084}[gr-qc]). 
\bib{WH-rev-3}
Lobo F. \textsl{arxiv:0710.4474}[gr-qc]. 
\bib{bronnikov-1}
Bronnikov B, Lemos J. \PRD{79}{2009}{104019} ~(\textsl{arxiv:0902.2360}[gr-qc]). 
\bib{bronnikov-2}
Bronnikov K, Skvortsova M, Starobinsky A. {\sl Grav. Cosmol.} {\bf 16},
(2010) 216 ~({\sl arxiv:1005.3262}[gr-qc]).
\bib{LC-BR-1}
T. Levi-Civita, \textsl{Rend. R. Acad. Naz. Lincei}, {\bf 26} (1917) 519.
\bib{LC-BR-2}
B. Bertotti, \PRD{116}{1959}{1331}.
\bib{LC-BR-3}
I. Robinson, \textsl{Bull. Akad. Pol.}, {\bf 7} (1959) 351.
\bib{eduardo-wh}
Guendelman E. Gen. Rel. Grav. \textbf{23} (1991) 1415. 
\bib{dzhunu-wh-1}
Dzhunushaliev V, Singleton D. \CQG{16}{1999}{973} ~({\sl arxiv:gr-qc/9805104}). 
%
\bib{dzhunu-wh-2}
Dzhunushaliev V, Singleton D. \PRD{59}{1999}{064018} ~({\sl arxiv:gr-qc/9807086}). 
%
\bib{dzhunu-wh-3}
Dzhunushaliev V, Kasper U, Singleton D. \PLB{479}{2000}{249} 
~({\sl arxiv:gr-qc/9910092}). 
%
\bib{dzhunu-wh-4}
Dzhunushaliev V. {\sl Gen. Rel. Grav.} \textbf{35} (2003) 1481 
~({\sl arxiv:gr-qc/0301046}). 
\bib{zaslavskii-mat}
Matyjasek J, Zaslavsky O. \PRD{64}{2001}{044005} ~({\sl arxiv:gr-qc/0006014}). 
%
\bib{zaslavskii-1}
Zaslavskii O. \PRD{70}{2004}{104017} ~({\sl arxiv:gr-qc/0410101}). 
%
\bib{zaslavskii-2}
Zaslavskii O. \PLB{634}{2006}{111} ~({\sl arxiv:gr-qc/0601017}). 
%
\bib{zaslavskii-3}
Zaslavskii O. \PRD{80}{2009}{064034} ~(\textsl{arxiv:0909.2270}[gr-qc]).
\bib{lindstroem-etal}
U. Lindstr\"{o}m and H. Svendsen, \IJMPA{16}{2001}{1347} 
(\textsl{arxiv:hep-th/0007101}).
\bib{inflation-all-1}
E. Guendelman, A. Kaganovich, E. Nissimov and S. Pacheva, 
\textsl{Centr. Europ. Journ. Phys.} {\bf 7} (2009) 668
(\textsl{arxiv:0711.2877}[hep-th]).
\bib{israel-poisson-1}
W. Israel and E. Poisson, \PRL{63}{1989}{1663}. 
\bib{israel-poisson-2}
W. Israel and E. Poisson, \PRD{41}{1990}{1796}.
\bib{randall-sundrum-1}
L. Randall and R. Sundrum, \PRL{83}{1999}{3370}
~(\textsl{arxiv:hep-ph/9905221]}).
%
\bib{randall-sundrum-2}
L. Randall and R. Sundrum, \PRL{83}{1999}{4690}
~(\textsl{arxiv:hep-ph/9906064]}).
%
\bib{mannheim}
P. Mannheim, {\em ``Brane-Localized Gravity''}, (World Scientific, 2005).
%
\bib{gregory}
R. Gregory, {\em Braneworld Black Holes}, ch.7 in \textsl{``Physics of Black
Holes: A Guided Tour''}, ed. by E. Papantonopoulos, \textsl{Lect. Notes Phys.}
{\bf 769}, pp.259-298 (Springer, Berlin Heidelberg 2009). 
%
\bib{maartens-koyama}
R. Maartens and K. Koyama, \textsl{Living Rev. Relativity} {\bf 13} (2010) 5
~(\textsl{arxiv:1004.3962}[hep-th]).
\bib{poisson-kit}
E. Poisson, {\em A Relativist's Toolkit. The Mathematics of Black-Hole
Mechanics}, (Cambridge Univ. Press, Cambridge UK, 2004).
\bib{rindler}
W. Rindler, {\em Relativity: Special, General and Cosmological}, 2nd ed.,
Sec.3.8 (Oxford Univ. Press, Oxford UK, 2006).
%
\bib{MTW}
Ch. Misner, K. Thorne and J. Wheeler, {\em Gravitation}, Sec.6.6
(W.H.Freeman and Co, USA, 1973).
\bib{eduardo-idan}
E. Guendelman and I. Shilon, \PRD{76}{2007}{025021} ~(\textsl{arxiv:gr-qc/0612057}).
\bib{misner-wheeler}
Ch. Misner and J. Wheeler, \AoP{2}{1957}{525-603}.
%
\bib{einstein-rosen}
A. Einstein and N. Rosen, \textsl{Phys. Rev.} {\bf 43}, 73 (1935).
%
\bib{Carroll}
S. Carroll, {\em ``Spacetime and Geometry. An Introduction to General Relativity''}
(Addison Wesley, San Francisco, 2003).
%
\bib{poplawski}
N. Poplawski, \PLB{687}{2010}{110} ~({\sl arxiv:0902.1994}[gr-qc]).
\bib{grav-cornell}
E. Guendelman, A. Kaganovich, E. Nissimov and S. Pacheva,
\PLB{704}{2011}{230}, erratum \PLB{705}{2011}{545} ~(\textsl{arxiv:1108.0160}[hep-th]).
\bib{hiding}
E. Guendelman, A. Kaganovich, E. Nissimov and S. Pacheva, 
\textsl{The Open Nuclear and Particle Physics Journal} {\bf 4} (2011) 27-34
(\textsl{arxiv:1108.3735}[hep-th]).
\bib{GG-1}
E. Guendelman, \IJMPA{19}{2004}{3255} ~(\textsl{arxiv:0306162}[hep-th]).
\bib{GG-2}
P. Gaete and E. Guendelman, \PLB{640}{2006}{201-204}
~(\textsl{arxiv:0607113}[hep-th]).
\bib{GG-3}
P. Gaete, E. Guendelman and E. Spalluci, \PLB{649}{2007}{217} 
~(\textsl{arxiv:0702067}[hep-th]).
\bib{GG-4}
E. Guendelman, \MPLA{22}{2007}{1209-1215} ~(\textsl{arxiv:0703139}[hep-th]).
\bib{GG-5}
I. Korover and E. Guendelman, \IJMPA{24}{2009}{1443-1456}.
\bib{GG-6}
E. Guendelman, \IJMPA{25}{2010}{4195-4220} ~(\textsl{arxiv:1005.1421}[hep-th]).
\bib{hide-confine}
E. Guendelman, A. Kaganovich, E. Nissimov and S. Pacheva, 
\textsl{arxiv:1109.0453}[hep-th], to appear in \textsl{Int. J. Mod. Phys.}.


\end{thebibliography}
\end{document}